\documentclass[11pt]{article}

\usepackage{mathrsfs}
\usepackage{epic, eepic}
\usepackage{subfigure}
\usepackage{amsfonts, amssymb, amsmath, graphicx, epsfig, lscape, capt-of}
\usepackage{url}
\usepackage{natbib}
\usepackage{ifpdf}
\usepackage[boxed]{algorithm2e}
\usepackage{color}

\def\inst#1{$^{#1}$}

\makeatletter

\begin{document}%


\title{Higher order assortativity for directed weighted networks and Markov chains}

\author{%
Alberto Arcagni\inst{1}\and Roy Cerqueti\inst{2,3}\and
Rosanna Grassi\inst{4,*}}
\date{}

\maketitle

\begin{center}
{\footnotesize
\vspace{0.2cm} \inst{1} MEMOTEF Department, Sapienza University of Rome, Via del Castro Laurenziano, 9 - 00161, Rome, Italy\\
\texttt{alberto.arcagni@uniroma1.it}\\
\vspace{0.3cm} \inst{2} Department of Economic and Social Sciences, Sapienza University of Rome, Piazzale Aldo Moro, 5 - 00185, Rome, Italy\\
\texttt{roy.cerqueti@uniroma1.it}\\
\vspace{0.3cm} 
\inst{3} GRANEM, Université d'Angers, SFR CONFLUENCES, F-49000 Angers, France
\\
\vspace{0.3cm} \inst{*} \textit{Corresponding author} \\ \inst{4} Department of Statistics and Quantitative Methods,  University of Milano - Bicocca, Via Bicocca degli Arcimboldi, 8 - 20126 Milan, Italy\\
\texttt{rosanna.grassi@unimib.it}}
\end{center}

\begin{abstract}

This paper proposes a new class of assortativity measures for weighted and directed networks. We extend the classical Newman's degree-degree assortativity by considering nodes' attributes different from the degree. Moreover, we propose connections among the nodes through directed paths of length greater than one -- thus, obtaining higher order assortativity. We provide an empirical application of these measures for the paradigmatic case of the trade network. Importantly, we show how this global network indicator is strongly related to the autocorrelations of the states of a Markov chain.


\vspace{5 mm} Keywords: Networks, Higher order Assortativity, Random walks, Markov Chains.
\end{abstract}

\section{Introduction}
Asymmetric interactions characterize many complex systems in nature.
Directed networks are suitable tools to represent these complex situations. Despite the remarkable persistence of this kind of systems in real-world phenomena, a formal representation of the directed networks is still not used, probably due to the difficulty in the mathematical treatment of such models. Indeed, some network indicators are not easy to manage in the directed framework; sometimes, and in some cases, a proper definition of the network measures for the directed networks is still missing.
Under this perspective, this paper focuses on the issue of providing a generalized concept of the assortativity measure for directed networks.
The theme is relevant because 
assortativity is a 
 global indicator that provides useful insights about the network structure. In the classical definition of \cite{Newman2002}, the assortativity is represented by a global measure based on
the Pearson correlation between the degrees of nodes. This measure ranges in the interval $[-1,1]$, having positive values in assortative networks and negative values in disassortative ones. The interpretation of the degree-degree assortativity measure is simple: 
undirected networks, high assortativity means that high-degree nodes tend to connect to other high-degree nodes. On the contrary, in disassortative networks,
high-degree nodes tend to be connected to low-degree
nodes.
The assortativity measure can be extended into two different 
directions. On the one side, one can consider other quantitative attributes of the nodes different from the degree; on the other side, one can move from the adjacency of the nodes -- which is the basis of Newman's degree-degree assortativity -- and propose more general ways to connect them. 
We move in the direction of meeting both such important requirements. 
Indeed, in the former case, degree-based assortativity provides a restrictive view of the accordance among nodes. A generalization to  different nodes' attributes overcomes this issue, extending the meaning of the coefficient to describe similarities/dissimilarities between nodes other than the degree (see \cite{Meghanathan2016}, \cite{arcagni2019extending}, \cite{Yuan2021}).
In the latter case, it is relevant to mention the generalization of Newman's assortativity index 
proposed in \cite{Arcagni2017}. The authors provide a closed formula that leads to a unifying approach of the assortativity for undirected and unweighted networks; in particular, they consider pairs of vertices not necessarily adjacent but also connected through paths, shortest paths and random walks. Importantly, connecting nodes through paths of length greater than one allows us to gain insights on the similarities between connected, but not adjacent nodes. Indeed, high-degree nodes may not be directly connected to each other, but remain equally connected through low-degree ones.
 
In \cite{Arcagni2017}, the new concept of assortativity coefficient based on paths is labelled as higher order assortativity.
The higher order assortativity allows us to recover, in particular cases, the Newman
index as well as other measures based on assortativity \citep{Estrada2008, Mayo2015, vanmieghen2010}, using suitable definitions of the matrix governing the connections.
In the same line, \cite{arcagni2019extending} extended the higher order assortativity to a new class of higher order measures specific for weighted undirected networks, recovering as a particular case the weighted assortativity index introduced in \cite{Leung2007}.
The problem of assortativity in directed networks has also been investigated \citep{Yuan2021}. In particular, \cite{Foster2010,Piraveenan2010} propose a natural modification of the Newman formula, encoding the assortativity measure into four directed measures on the basis of the combination of nodes' in- and out-degrees. 
The coefficients quantify the tendency of nodes with high in-degree to be connected to nodes with high in-degree -- and the same applies to the other cases in-degree with out-degree, out-degree with in-degree and out-degree with out-degree. However, a higher order measure of assortativity tailored for weighted and direct networks is still missing.

This paper contributes to this debate by filling this gap. We introduce a new concept of assortativity measure of order $h$ -- with $h \geq 1$ -- for the general case of weighted and directed networks. According to the previous literature and when needed, we briefly refer to such a concept as the higher order assortativity measure. We here complete the definitions introduced in \cite{Arcagni2017, arcagni2019extending} by including also the weighted directed networks. Moreover, the proposed concept is formulated on a node attribute that is not necessarily the degree or the strength. In doing so, we extend the Newman assortativity coefficient in both the directions described above.



It is important to notice that the proposed assortativity measure offers a relevant interpretation in the field of the Markov chains \citep[see, e.g.,][for a survey on this class of stochastic processes]{norris}.
Indeed, the directed and weighted networks are particularly suitable for the description of random moving flows over a discrete set of states identified by nodes. In the graph representation of a Markov chain, the existence of an arc connecting state/node $i$ 
with state/node $j$ is associated with a non-null probability of 
jumping from $i$ to $j$. Therefore, the (suitably normalized) weights of the arcs in a complex network are naturally associated with the probability to observe a flow from a state to another one in the corresponding Markov chain. 
In stating the bridge between complex networks and Markov chains, we follow the route traced by some relevant contributions in the literature. It is worth mentioning \cite{gomez}, where the authors deal with a disease propagation model by merging complex networks and Markov chains. In the same application environment, \cite{iannelli} elaborate on the informative content of the general network-based measures derived from Markov chain theory.

Our novelty here is to connect the higher order assortativity -- a topological property of the network involving nodes beyond the adjacent ones -- and homogeneous discrete-time Markov chains.


Specifically, let us consider the nodes' assortativity 
referred to a given vertex centrality measure. The assortativity index of order $h$ ranges in $[-1,1]$; it approaches $1$ ($-1$, resp.) when the nodes connected through walks of length $h$ have a high level of concordance (discordance, respectively) in the considered centrality measure. Such a variation range suggests that the assortativity index of order $h$ can be interpreted as an autocorrelation term at $h$ lags in the context of Markov chains. This is exactly our case. Indeed,
the order $h$ is the length of a walk from one node to another. If we imagine that each link takes one unit of time, then the considered walk takes $h$ units of time. In this respect, the order $h$ can be seen as a temporal variable for a discrete-time Markov chain with states given by the nodes of the network along with their centrality measures and transition probabilities opportunely related to the weights of the network's links. 

This interpretation of the assortativity measure of order $h$ allows stating a natural bridge between complex networks and stochastic processes. In so doing, we are able to move from the information content of the higher order assortativity of a network to the dynamical properties of the underlying Markov chain. Specifically, the temporal dimension of the network and the regularities captured by the autocorrelation – which are hidden in the network structure – become clear in moving to the Markov chain theory.

The paper is structured as follows. In Section \ref{sec:preliminaries} we describe the mathematical notations. In Section \ref{Sec. HOAss} we introduce  a novel definition of higher order assortativity extended to weighted digraphs. In Section \ref{Sec. ApplWTN} we show an application of the proposed index to the trade network. Section \ref{sec5} illustrates how the higher-order assortativity measure can be interpreted in the context of the discrete-time homogeneous Markov chains with finite states. Conclusions follow in Section \ref{Sec. Conc}. 

\section{Preliminaries}\label{sec:preliminaries}
In this section we outline of some basic definitions and notation related to directed complex networks 
that are used to present our methodological proposal.



For a more detailed treatment 
we refer, for instance, to \cite{Bang2008, Newman2010_DOPPIO,Gross2003}.

A directed graph (or digraph) $N=(V,A)$ is a pair of sets $V$ and $A$, where $V$ is the set of $n$ vertices (or nodes) and $A$ is the set of $m$ ordered pairs (arcs) of vertices of $V$; if $(i,j)$ and/or $(j,i)$ is an element of $A$, then vertices $i$ and $j$ are adjacent.
A $i-j$ directed walk is a sequence of
vertices and arcs from $i$ to $j$ such that all arcs have the same direction.

A weight $w_{ij}>0$ can be associated with each arc $(i,j)$ so that a weighted digraph is obtained. Moreover, it is assumed that $w_{ij}=0$ if and only if $(i,j) \notin A$. The weights $w$'s are collected in a real $n$-square matrix $\textbf{W}$ (the weighted adjacency matrix). By definition, the elements of such a matrix describe both the adjacency relationships between the vertices in $V$ and the weights on the arcs. 
In this context, a weighted directed network is a directed graph $N=(V,A)$ equipped by a weighted not symmetric matrix $\textbf{W}$.
In the unweighted case, non-null weights can take only unitary value, so that matrix $\textbf{W}$ provides only information about the adjacency relationships. In the case of unweighted networks, we denote $\textbf{W}$ by $\textbf{A}$ (the adjacency matrix).

The $(i,j)$ element of the $k-$power of the matrix \textbf{A} is the number of directed walks of length $k$ from $i$ to $j$. 


Since in-degree, out-degree, in-strength and out-strength are the most popular centrality measures in the context of the assortativity -- as the original definition of \cite{Newman2002} suggests -- we introduce a particular notation for them, along with their definition.

We define the in-degree $d^{I}_i$ (respectively out-degree $d^{O}_i$) of a node $i \in V$ as the number of arcs pointing towards (respectively starting from) $i$. Then, the in-degree and out-degree vectors are, respectively $\mathbf{d}^{I}$, and $\mathbf{d}^{O}$.
The vector of the vertices' degrees is given by $\mathbf{d}=\mathbf{d}^{I}+\mathbf{d}^{O}$.
Analogously, definitions of the in-strength and out-strength of node $i \in V$ ($s^{I}_i$ and $s^{O}_i$, respectively) and in-strength, out-strength vectors ($\mathbf{s}^{I}$ and $\mathbf{s}^{O}$, respectively) are provided.
A vertex $i \in V$ with $d^I_i = 0$ and $d^O_i > 0$ is called \emph{source}, as it is the origin of its outcoming arcs. Similarly, a vertex $i \in V$ with $d^O_i = 0$ and $d^I_i > 0$ is called \emph{sink} since it is the end of each incoming arc.

\section{A novel definition of higher order assortativity}\label{Sec. HOAss}

We consider two centrality measures for the vertices of a weighted directed network and collect their value the definition of higher order assortativity for weighted digraphs, thus extending the related concepts introduced by \cite{Arcagni2017,arcagni2019extending} as a particular application of the Pearson correlation index in the $n$-dimensional vectors $\textbf{x}$ and $\textbf{y}$. 
The higher order assortativity index of length $h\geq 1$ for a directed graph and associated with $\textbf{x}$ and $\textbf{y}$ can be defined as follows
\begin{equation}
  r_h=r(\textbf{x}, \textbf{y},\mathbf{E}_h) = \frac{
    \mathbf{x}^\top\left(\mathbf{E}_h - \mathbf{p}_h\mathbf{q}_h^\top\right)\mathbf{y}
  }{
    \sqrt{\left[
      \mathbf{x}^\top\left(\mathbf{D_p}_h - \mathbf{p}_h\mathbf{p}_h^\top\right)\mathbf{x}
    \right]\left[
      \mathbf{y}^\top\left(\mathbf{D_q}_h - \mathbf{q}_h\mathbf{q}_h^\top\right)\mathbf{y}
    \right]},
  }
  \label{eq:higher}
\end{equation}
where $\textbf{E}_h$ is a normalized $n$-squared matrix whose element on the $i$-th row and $j$-th column is denoted by $e^{(h)}_{ij}$, with $\Vert\textbf{E}_h\Vert_1=\sum_{i,j=1}^{n}\left|e_{ij}\right| =1$ 
while $\textbf{p}_h = \textbf{E}_h\textbf{1}$, $\textbf{q}_h = \textbf{E}_h^\top\textbf{1}$, where $\textbf{1}$ is the conformable vector of elements equal to 1, and $\mathbf{D_p}_h$ and $\mathbf{D_q}_h$ are the diagonal matrices with the vectors $\textbf{p}_h$ and $\textbf{q}_h$, respectively, on the main diagonal.

In particular we assume that  $\textbf{E}_1 = \frac{\textbf{W}}{\Vert\textbf{W}\Vert_1}$ and, 
as we will see in Section \ref{sec:hoaMc}, matrix $\textbf{E}_h$ is also related to Markov chains $ \forall \,h \geq 1$. Therefore, the assortativity measure in equation \eqref{eq:higher} depends on suitable centrality measures $\textbf{x}$ and $\textbf{y}$, that are related to the initial and terminal state, respectively, of a stochastic process of $h$ steps.


Accordingly to \cite{Newman2002}, we point out that the in-strength and out-strength of the nodes plays a relevant role when dealing with assortativity also in the context of weighted digraphs. 
Indeed, such centrality measures provide distinct information about the network topology.
Therefore, we describe such cases in detail and -- at the same time -- we also give the interpretation of the higher-order assortativity measure in this particular context.


Specifically, for a complete treatment of the higher-order assortativity measure for weighted and directed networks, we consider all the possible combinations of in-strength and the out-strength and, starting from equation \eqref{eq:higher}, we define the four higher-order assortativity indexes for weighted digraph. 

\begin{enumerate}
	\item  $r^{OI}_h = r(\textbf{s}^O, \textbf{s}^I,\mathbf{E}_h)$ (out-in)\\
	 This coefficient evaluates the correlation between the out-strength and in-strength vectors using formula (\ref{eq:higher}), for nodes that are connected through a directed walk of length $h$. Specifically, a high value of $r^{OI}_h$ is associated to high concordance between the out-strength and the in-strength of such vertices, while a value of $r^{OI}_h$ close to -1 relies to high discordance. 

   \item $r^{IO}_h = r(\textbf{s}^I, \textbf{s}^O,\mathbf{E}_h)$ (in-out)\\ 
   This index measures the concordance and discordance between the in-strength of the starting vertices with the out-strength of the arrival vertices after a directed walk of length $h$.
    \item   $r^{II}_h= r(\textbf{s}^I, \textbf{s}^I,\mathbf{E}_h)$ (in-in) \\
    This coefficient provides an evaluation of the correlation between the in-strength of the starting vertices with the in-strength of the ending vertices in presence of a directed walk of length $h$.

  \item  $r^{OO}_h = r(\textbf{s}^O, \textbf{s}^O,\mathbf{E}_h)$ (out-out) \\ This higher order assortativity index captures the correlation between the out-strength of the starting vertices with the out-strength of the terminal ones when they are connected through a directed walk of length $h$.
 
\end{enumerate}


Notice that, given $h$, not necessarily $r^{OI}_h = r^{IO}_h$ because the indexes depend the distribution of directed walks described by the matrix $\textbf{E}_h$ that is not necessarily symmetric.

\section{Higher order assortativity in the world trade network}\label{Sec. ApplWTN}
To illustrate the advantages of our definition we show an application of the higher order assortativity to the trade network.\\
Data refer to traded goods between countries, the volume of trades are expressed in US dollars\footnote{Data are free and available for the download by  ComTrade - see \cite{UNComtrade}}.
We constructed both export and import weighted networks where vertices are countries and weights on arcs represent the volume of imported or exported goods between 222 countries in the year 2021. 
Notice that the number of vertices is greater than the recognized countries, because ComTrade includes even almost uninhabited territories.
Therefore, in the following, the word ``country'' covers a broad definition that also includes these regions.

Concerning links between countries, the export network is expected to be complementary to the import network, as an import from $i$ to $j$ should be equal to the export from $j$ to $i$. 
This is not the case because data describe trades officially declared by \emph{reporters} countries with their \emph{partners}, and some declarations are missed or mismatch. This problem still persists also considering years before 2021.

Here we followed an approach similar to the one adopted by \cite{antonietti2022world}, that constructed a symmetric trade network aggregating import and export data. However, unlike \cite{antonietti2022world}, in our network we preserve the arc directions, working in this way with a directed network.

More specifically, let $\mathbf{W}_I$ and $\mathbf{W}_E$ be weighted adjacency matrices representing the volume of import and export trades among countries. Thus, the trade network is defined as $N = (V, \mathbf{W})$ where the elements $w_{ij}$ of the matrix $\mathbf{W}$ are:
$$
w_{ij} = \max(w_{ij}^E, w_{ji}^I),
$$
being $w_{ij}^E$ and $w_{ji}^I$ the $ij-$ entry of matrices $\mathbf{W}_E$ and $\mathbf{W}_I^\top$, respectively. In this way, the asymmetric matrix $\mathbf{W}$ expresses the flow of goods from country $i$ to $j$. It is worth noting that in constructing $\mathbf{W}$ we choose the maximum between the import and export. Indeed, in line with 
\cite{antonietti2022world}, we assume that the higher value declared, the higher the quality of the information that can be contained.

\subsection{Analysis of the main characteristics of the network}
At first, we analyse some classical network characteristics.

The number of vertices is $n = 222$.
Let $\mathbf{A}$ be the adjacency matrix obtained by $\mathbf{W}$ by neglecting arcs' weights.
The network \emph{density} is computed as the ratio between the total number of arcs and the maximum number of potential arcs\footnote{$N$ is a directed network without self-loops} $(n - 1)n$.

Therefore, the density is
$$
\frac{\Vert\textbf{A}\Vert_{1}}{(n - 1)n} =\frac{\sum_{i,j=1}^{n}a_{ij}}{(n - 1)n} = 0.4996.
$$

Although one might expect the trade network to be complete, this is not true in this case.
This is due to the presence of small countries and regions representing peripheral nodes, that probably communicate only a part of their exports or imports.

\begin{figure}[!h]
\begin{center}
\includegraphics[width = \textwidth]{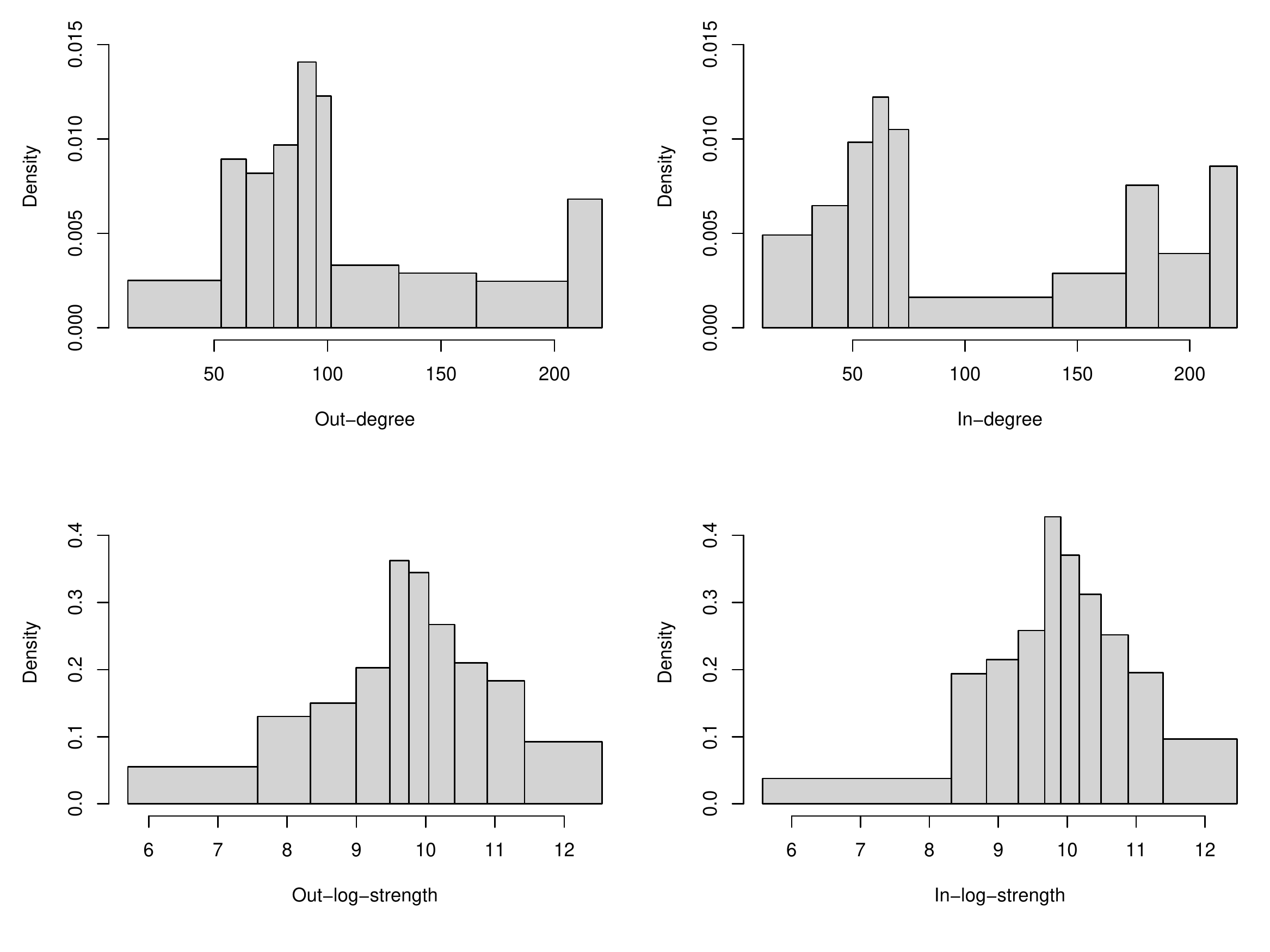}
\end{center}
\caption{Marginal distributions of the considered centrality measures. Histogram classes represent the deciles.}
\label{fg:marginal}
\end{figure}

Distributions of centrality measures, in particular degree and strength, provide other important information. 
The degree and strength of a vertex represent the number and volume 
of a country's exchanges, then we get insight of the diversification and intensity of trades. 
Figure \ref{fg:marginal} shows the marginal distributions of in-out degree and in-out strength. In- and out-strength are graphically represented using the logarithmic scale with base 10.

The distributions of ``in'' and ``out'' measures are quite similar, but it is worth stressing some characteristics of the degree and log-strength distributions:
\begin{enumerate}
\item on the $x-$ axis, the magnitude of the degrees corresponding to the number of countries is few hundreds, whereas that of the strength ranges from $10^6$ to $10^{12}$;
\item the log-strength distribution is bell-shaped;
\item the degree distribution is polarized, as more than half of the countries have connections with less than 100 other countries, and one-tenth of the countries are connected with almost all the other ones.
\end{enumerate}
As we will see later, this last characteristic is meaningful in searching for clusters of central and peripheral countries.

We now compare countries through the four centrality measures: in-degree, out-degree, in-log-strength and out-log-strength.

Table \ref{tb:corr} shows the strong correlation (computed using the Pearson coefficient) between the considered variables.

\begin{table}
\centering
\begin{tabular}{l|llll}
& in-deg. & out-deg. & in-log-str. & out-log-str. \\ \hline
in-deg. & 1 & & & \\
out-deg. & 0.883 & 1 & & \\
in-log-str. & 0.792 & 0.785 & 1 & \\
out-log-str. & 0.742 & 0.771 & 0.947 & 1
\end{tabular}
\caption{Correlation matrix between centrality measures.}
\label{tb:corr}
\end{table}
As expected, ``in'' and ``out'' directions are strongly correlated ($0.883$ and $0.947$ for degree and log-strength, respectively). However, there is also a strong \emph{linear} correlation between degree and log-strength centralities with values ranging from $0.771$ to $0.792$.
Although the correlation is not causation, these high values suggest that an increasing number of connections increases volumes exponentially.

We now analyse the principal components from the standardized variables.
The first component represents 87\% of the whole variability (considering the first two together, the variability rises to 96\%), and its loading is all positive. Therefore, its scores can be used to get a mixed and rough order of the countries in terms of centrality.
The top 10 of the central countries are, in increasing order, Canada, Belgium, Japan, Italy, Great Britain, Netherlands, France, Germany, China and the United States.

We performed a hierarchical cluster analysis using the four principal components instead of the strongly correlated original variables.
We used Euclidean distances between countries, and we defined distances between groups with the complete method.

The number of groups suggested by the hierarchical method is two.
One cluster of 116 is composed of small territories and countries almost peripheral in terms of volumes, and the other cluster of the remaining 106 countries controls most of the volumes. We refer to this cluster as ``main network''.\\
\begin{figure}[!h]
\begin{center}
\includegraphics[width=\textwidth]{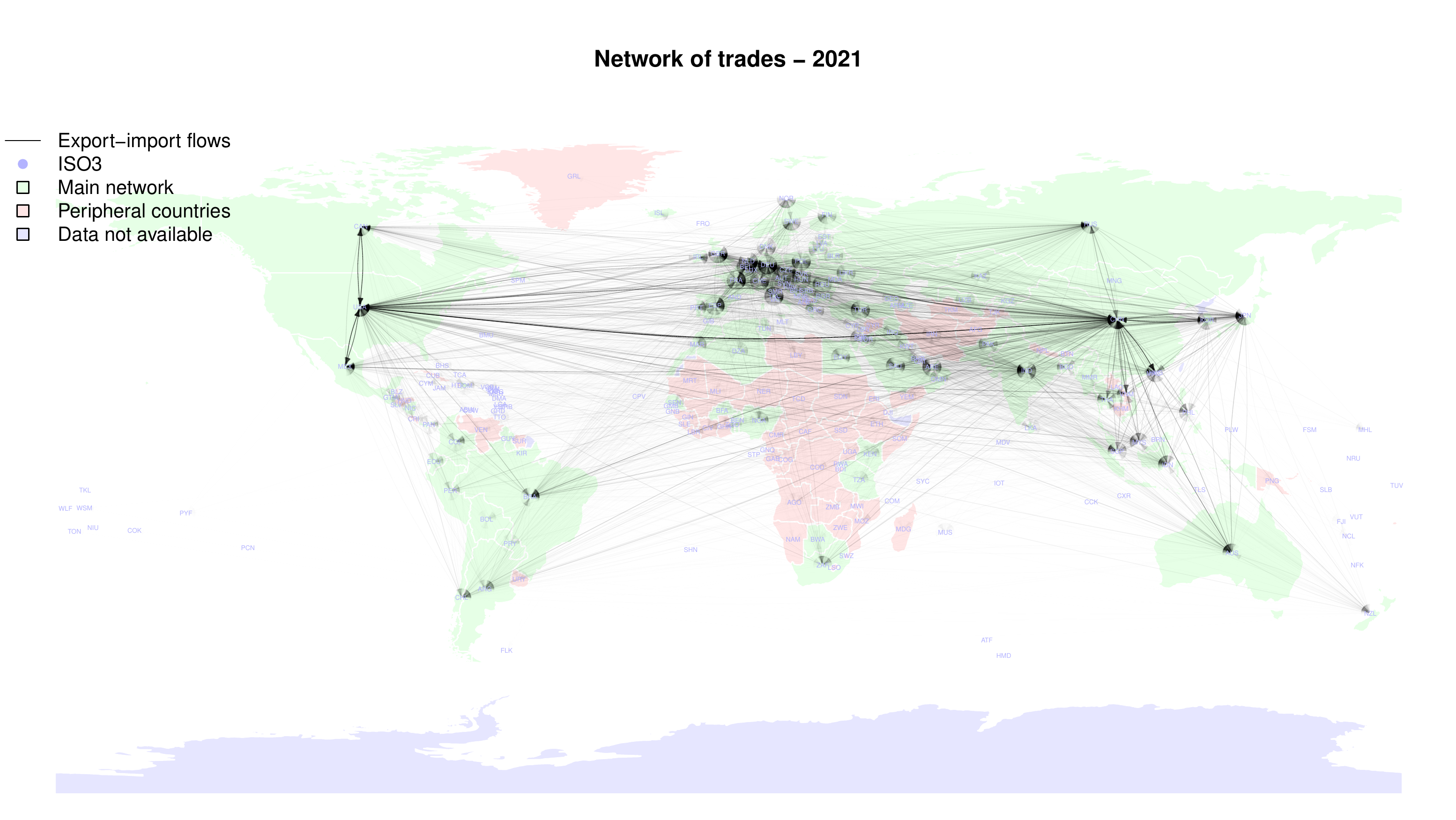}
\end{center}
\caption{World map of the trade network. Countries are coloured in green and pink, according to the cluster they belong to.}
\label{fg:map_clust}
\end{figure}
Figure \ref{fg:map_clust} provides a graphic representation of the trade network, showing the groups that emerged from the cluster analysis, namely the main network and peripheral countries.\\ However, how these groups relate to each other is a question that still needs to be investigated. As we will see in the next subsection, assortativity provides an answer in terms of centrality correlation.

\subsection{Higher order assortativity}

Evidently, there are several combinations of centrality measures and adjacency matrices to be used for computing the higher order assortativity coefficient. In this section, we focus only on the cases offering a more intuitive interpretation of the results in the context of world trade\footnote{We also computed the higher order assortativity measures in all the remaining cases and results are available upon request.}.


It is important to observe that, working on  directed networks, in the assortativity coefficient defined in formula \eqref{eq:higher} there is not symmetry between vectors $\textbf{x}$ and $\textbf{y}$. Indeed, in this application  $\textbf{x}$ and $\textbf{y}$ are centrality measures and, as pointed out in Section \ref{Sec. HOAss}, the resulting coefficient depends on the out and in-centralities we stand for $\textbf{x}$ and $\textbf{y}$. We give the following economical interpretation of the used centrality measures:
\begin{itemize}
\item the out-going centrality measures the supply, i.e., out-degree and out-strength of vertex $i$ is the supply of good of country $i$.
\item the in-going centrality measures the demand, i.e., in-degree and in-strength of vertex $i$ is the demand of good of country $i$.
\item the direction of the edges represents the export,  i.e., an arrow from $i$ to $j$ represents the export from country $i$ to country $j$.
\end{itemize}

In view of the economic meaning of centrality measures given above, 
the most suitable directed assortativity in this application is the out-in assortativity $r^{OI}_h$ 
in order to effectively track the paths of goods' exchanges between countries. 



\begin{figure}[!h]
\begin{center}
\includegraphics[width=0.45\textwidth, page = 1]{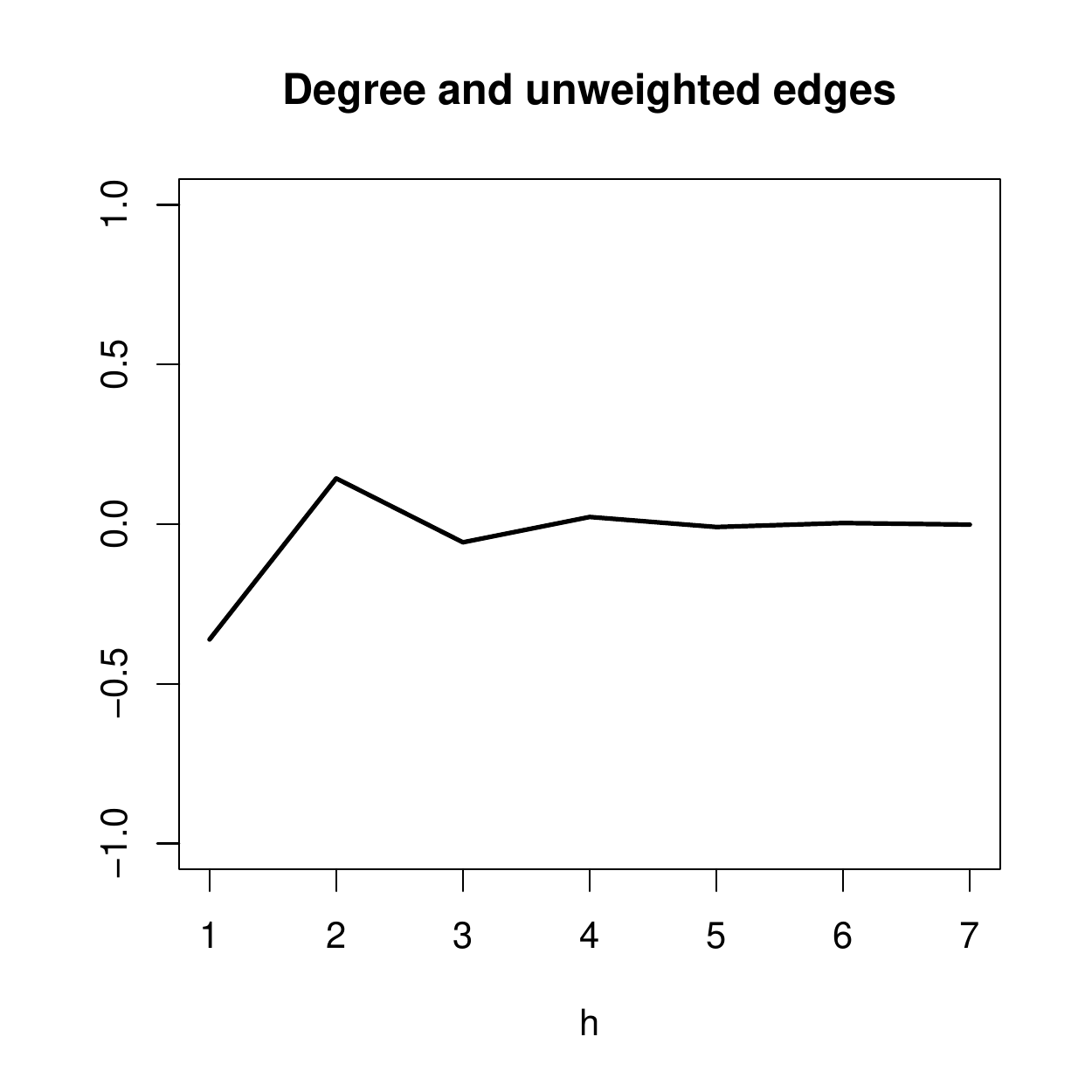}
\includegraphics[width=0.45\textwidth, page = 6]{final_results}
\end{center}
\caption{Export-import higher order assortativity for the unweighted (left) and weighted (right) network.}
\label{fg:higherOrderAssortativity}
\end{figure}

Figure \ref{fg:higherOrderAssortativity} depicts the higher order direct assortativity for increasing $h$ computed on both the unweighted and weighted network.
In both cases, the network is disassortative at order $h = 1$; this means that the exporters with a large number of connections and a large volume tend to trade with small importers.
Note that, in absolute value, the assortativity is immediately lower than $0.5$, and it is due to big traders that also need to trade one each other (for instance, The United States and China) and maybe also small traders.\\
At order $h = 2$ the assortativity becomes positive. A similar behaviour has been also observed for undirected networks \citep[see][]{Arcagni2017, arcagni2019extending}, and it may be the consequence of the disassortative behaviour at order $h = 1$.
Indeed, if, at the first step, exporters and importers tend to belong to different groups, then in step two, they tend to belong to the same one.
This bounce between disassortative and assortative behaviour continues for odd and even orders, respectively.

Considering the weights the behaviour of higher order assortativity is similar because of the correlations between centrality measures reported in Table \ref{tb:corr}. 
However, considering weights, we observe a strong reduction in the amplitude of variation.
In fact, weights on arcs separate the main network, identified with the cluster analysis, from the remaining, and the main network of trades is almost complete as expected.

Finally, we observe that assortativity tends to $0$ when the order increases because of the increment of randomness in the expected flow of goods in the network.

\section{A Markov chains-based interpretation of the higher order assortativity}
\label{sec5}
In this section, we formalize the informative content of the higher-order assortativity measure in terms of the transition probabilities of discrete-time homogeneous Markov chains with finite states. 
As a premise, we introduce some basic definitions and notations for the Markov chains.
For a complete treatment, the reader can refer, for instance, to \cite{norris}.

\subsection{Markov chains: basic definitions and notation}

A homogeneous discrete-time Markov chain with finite states is a stochastic process $\mathcal{X}=(X(t):t \in \mathbb{N})$, such that the following conditions are true.
\begin{itemize}
\item[$(P1)$] There exists a set $V$ with $n$ elements such that $X(t) \in V$, for each $t \in \mathbb{N}$. The elements of $V$ are the states of $\mathcal{X}$, and the set $V$ is the states space of the Markov chain.
\item[$(P2)$] The Markov property holds, i.e.:
$$
P(X(t+1) = j | X(t)=i, X(t-1)=i_{t-1}, \dots, X(0)=i_0) = P(X(t+1) = j | X(t)=i),
$$
for each $t \in \mathbb{N}$ and $i,j, i_{0}, \dots, i_{t-1} \in V$.
\item[$(P3)$] The one-step transition probabilities are invariant with respect to time, i.e.:
$$
P(X(t+1) = j | X(t)=i) = P(X(1) = j | X(0)=i),
$$
for each $t \in \mathbb{N}$ and $i,j \in V$. We collect all the one-step transition probabilities in a $n$-square matrix $\mathbf{P}$, whose generic element is $p_{ij}=P(X(1) = j | X(0)=i)$. $\mathbf{P}$ is the transition probability matrix of the Markov chain $\mathcal{X}$.
\end{itemize}

As we will see below, the nodes set $V$ of the network coincides with the states space of the Markov chain.

Properties $(P1)-(P3)$ allow us to identify a Markov chain through three elements: the states space $V$, the transition probability matrix $\mathbf{P}$ and the initial probability distribution $\pi_0=(\pi_0(1), \dots, \pi_0(n))$, where $\pi_0(i)=P(X(0)=i)$, for each $i \in V$. In particular, condition $(P3)$ states that the Markov chain $\mathcal{X}$ is homogeneous.

Furthermore, property $(P3)$ can be extended to the case of $h$-steps transition probabilities that are invariant with respect to time for each $h \in \mathbb{N}$. A straightforward computation shows that
\begin{equation}
\label{pij^l}
P(X(t+h) = j | X(t)=i) = P(X(h) = j | X(0)=i),
\end{equation}
for each $t,h \in \mathbb{N}$ and $i,j \in V$. We collect the $h$-steps transition probabilities in the $n$-square matrix $\mathbf{P}_{h}$. The element of such a matrix in the $i$-th row and $j$-th column is $p^{(h)}_{ij}=P(X(h) = j | X(0)=i)$.

Since the states space $V$ is finite, the matrix $\mathbf{P}_{h}$ is $\mathbf{P}^h$, i.e., the $h$-steps transition probability matrix is the $h$-th power of the one-step matrix $\mathbf{P}$.

\subsection{Higher-order assortativity in the context of Markov chains}
\label{sec:hoaMc}

Given $h \geq 1$, we recall that $\Vert\textbf{E}_h\Vert_1=1$; therefore, by referring to the Markov chains theory, the generic element $e^{(h)}_{ij}$ may represent the joint probability that $i$ and $j$ are the position states of $\mathcal{X}$ at time $t$ and $t+h$, respectively, for each $t$, i.e.:
\begin{equation}
e_{ij}^{(h)}=
P(X(h) = j , X(0)=i).
\label{eq:eij^h}
\end{equation}

By considering the Markov chain $\mathcal{X}$ for which (\ref{eq:eij^h}) is true and by using the conditioned probability definition, we can write
\begin{equation}
\mathbf{E}_h = \mathbf{D}_{\pi_0} \mathbf{P}^h
\label{eq:joint_l}
\end{equation}
where $\mathbf{D}_{\pi_0}$ is a diagonal $n$-squared matrix whose diagonal entries are the components of the initial distribution $\pi_0$.
%
%
Being matrix $\mathbf{P}$, and then $\mathbf{P}^h$ with $h>1$, stochastic by row, $\mathbf{P}^h\textbf{1}=\textbf{1}$, hence:

\begin{equation}
\label{initial_prob_distr}
\textbf{E}_h\textbf{1}=(\mathbf{D}_{\pi_0}\mathbf{P}^h)\textbf{1}=\mathbf{D}_{\pi_0}(\mathbf{P}^h\textbf{1})=\mathbf{D}_{\pi_0}\textbf{1}=\pi_0
\end{equation}
then $\pi_0=\textbf{p}_h$, $h \geq 1$, where $\textbf{p}_h$ is the vector appearing in formula \eqref{eq:higher}.
Indeed, being $e^{(h)}_{ij}$ joint probabilities, by equation \eqref{eq:eij^h}:
$$
\sum_{j=1}^{n}e_{ij}^{(h)}=\sum_{j=1}^{n}P(X(h) = j , X(0)=i)=P(X(0) = i)=\pi_0(i)\, \forall i=1,...,n
$$
that is, $\textbf{p}_h$ is the initial probability distribution.

It is clear that, in general, there is not a unique Markov chain $\mathcal{X}$ with state space $V$ for which (\ref{eq:joint_l}) is valid.
However, having set $\textbf{E}_1=\frac{\textbf{W}}{\Vert\textbf{W}\Vert_1}$, in virtue of equation \eqref{initial_prob_distr}, there is a unique vector $\pi_0$.
So, once we choose the network $N=(V,A)$, there is a unique Markov chain $\mathcal{X}=(X(t):t \in \mathbb{N})$, with $V,\mathbf{P}$ and $\pi_0$ fixed.

Note that for each node $i \in V$ that in the network is a sink, matrix $\mathbf{P}$ has to be stochastic; therefore, a $1$ is added in the main diagonal to associate to the sink an absorbing state in the Markov chain, and $\pi_0(i)$ has to be $0$ to do not modify the topology of the network and comply with equation \eqref{eq:joint_l} for $h = 1$.
On the contrary, $\pi_0(i) > 0\,\forall i \in V$ if $i$ is not a sink because \eqref{eq:joint_l} returns a zero line in $\textbf{E}_1$ no matter the values in $\textbf{P}$.

Analogously to equation \eqref{initial_prob_distr},
\begin{equation}
\label{final_prob_distr}
\textbf{E}_h^\top\textbf{1}=(\mathbf{D}_{\pi_0}\mathbf{P}^h)^\top\textbf{1}=(\mathbf{P}^h)^\top\mathbf{D}_{\pi_0}\textbf{1}=(\mathbf{P}^h)^\top\pi_0=\pi_h
\end{equation}
then $\pi_h=\textbf{q}_h$, $h \geq 1$, where $\textbf{q}_h$ is the vector appearing in formula \eqref{eq:higher}. Again, by eq. \eqref{eq:eij^h}:
$$
\sum_{i=1}^{n}e_{ij}^{(h)}=\sum_{i=1}^{n}P(X(h) = j , X(0)=i)=P(X(h) = j)=\pi_h(j)\, \forall j=1,...,n
$$
that is, $\textbf{q}_h$ is the probability distribution after $h$ steps.

Keeping this in mind, we can rewrite the higher order assortativity index (defined in equation \ref{eq:higher}) in Markov chains context:

\begin{equation}
r_h=r(\textbf{x}, \textbf{y},\mathbf{D}_{\pi_0} \mathbf{P}^h) = \frac{
\mathbf{x}^\top\left(\mathbf{D}_{\pi_0} \mathbf{P}^h - \pi_0\pi_h^\top\right)\mathbf{y}
}{
\sqrt{\left[
\mathbf{x}^\top\left(\mathbf{D}_{\pi_0} -
\pi_0\pi_0^\top\right)\mathbf{x}
\right]\left[
\mathbf{y}^\top\left(\mathbf{D}_{\pi_h} -
\pi_h\pi_h^\top\right)\mathbf{y}
\right]},
}
\label{eq:mc}
\end{equation}

Then, the higher order assortativity index of length $h$ represents the autocorrelation at time-distance $h$ of a Markov chain, not evaluated on the states $v_i \in V$ but on their transformations $\mathbf{x}$ and $\mathbf{y}$.
Indeed, in a network context, centrality measures can be the image of a real function on the node set $V$. 

\subsection{Some illustrative examples}
We present some toy examples to illustrate how the newly introduced assortativity measure works in the context of Markov chains.
To this aim, we consider some instances of weighted networks whose related Markov chains have a meaningful state classification.
We enter the details.

\begin{figure}[h!]
\begin{center}
\includegraphics[width = 0.4\textwidth, page=1]{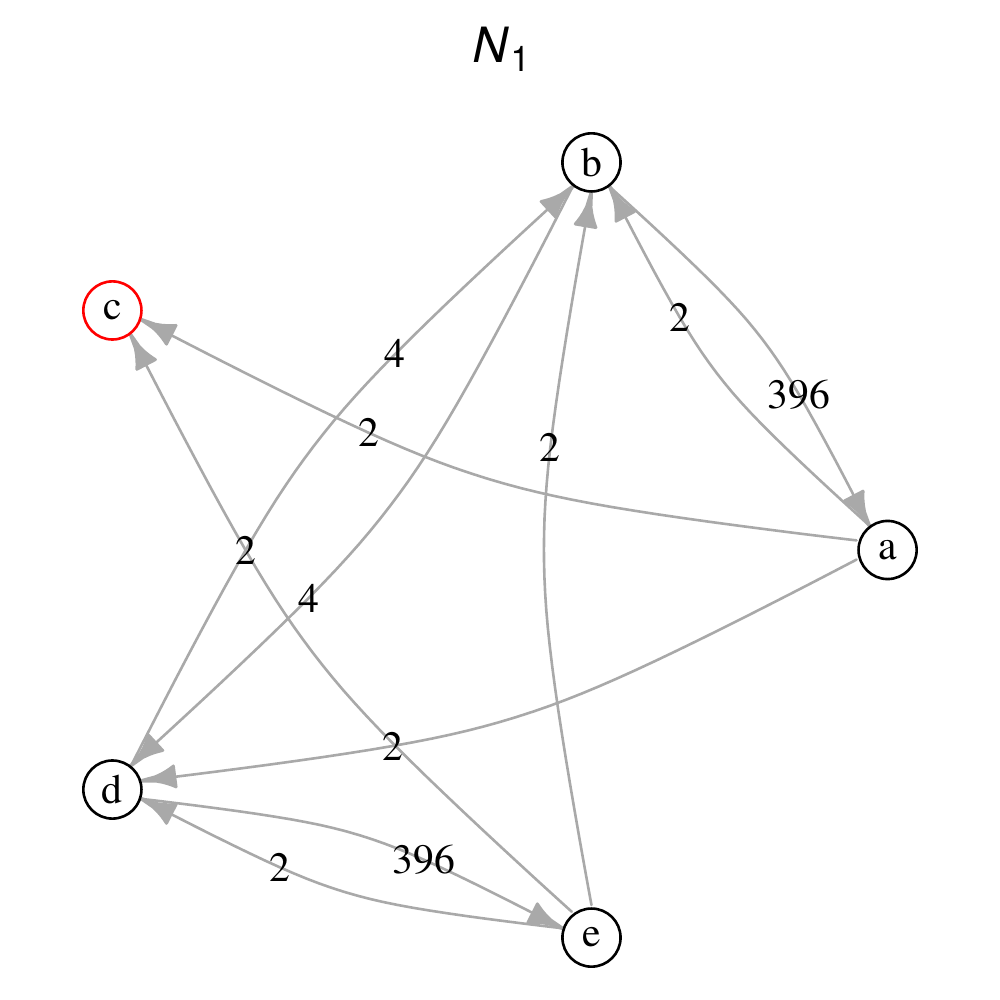}
\includegraphics[width = 0.4\textwidth, page=2]{toy}
\includegraphics[width = 0.4\textwidth, page=3]{toy}
\includegraphics[width = 0.4\textwidth, page=4]{toy}
\end{center}
\caption{Graphical representations of the four illustrative examples.}
\label{fg:examples}
\end{figure}

%
We consider four networks
$N_k=(V, \textbf{W}_k)$, with $k=1, \dots, 4$, sharing the same set of nodes $V=\{a,b,c,d,e\}$, and with weighted adjacency matrices $\textbf{W}_k$ defined as follows:
$$
\textbf{W}_1=\left(
\begin{array}{ccccc}
0 & 2 & \mathbf{2} & 2 & 0 \\
396 & 0 & 0 & 4 & 0 \\
0 & 0 & 0 & 0 & 0 \\
0 & 4 & 0 & 0 & 396 \\
0 & 2 & \mathbf{2} & 2 & 0 \\
\end{array}
\right),\qquad
\textbf{W}_2=\left(
\begin{array}{ccccc}
0 & 2 & \mathbf{196} & 2 & 0 \\
396 & 0 & 0 & 4 & 0 \\
0 & 0 & 0 & 0 & 0 \\
0 & 4 & 0 & 0 & 396 \\
0 & 2 & \mathbf{196} & 2 & 0 \\
\end{array}
\right)
$$
$$
\textbf{W}_3=\left(
\begin{array}{ccccc}
0 & \mathbf{2} & 198 & 0 & 0 \\
0 & 0 & 0 & 100 & 0 \\
150 & 0 & 0 & 0 & 150 \\
0 & 200 & 0 & 0 & 0 \\
200 & 0 & 200 & 0 & 0 \\
\end{array}
\right),\qquad
\textbf{W}_4=\left(
\begin{array}{ccccc}
0 & \mathbf{198} & 198 & 0 & 0 \\
0 & 0 & 0 & 100 & 0 \\
150 & 0 & 0 & 0 & 150 \\
0 & 200 & 0 & 0 & 0 \\
200 & 0 & 200 & 0 & 0 \\
\end{array}
\right)
$$
The networks are shown in Figure \ref{fg:examples}.
By inspection of the figure, it is worth noting that $N_1$ and $N_2$ have the same topology and differ only by two weights.
In particular, there are no outflows from $c$; that is, $c$ is a sink.
However, while in the former case the inflows in $c$ have remarkably low weights, in the latter, the inflows in $c$ are particularly high.
Such evidence is true in terms of weights' absolute values (i.e. considering matrices $\textbf{W}_1$ and $\textbf{W}_2$) but also under a relative perspective (i.e. considering the transition probability matrices $\textbf{P}_1$ and $\textbf{P}_2$ shown below).

Also, $N_3$ and $N_4$ share the same topology, and there are no outflows from the class of nodes $\{b,d\} \subset V$.
In both cases, it is possible to have an inflow in such a class from the set $\{a,c,e\}$, through the arc from $a$ to $b$.
Such an arc has a low weight in network $N_3$ and remarkably high for $N_4$.


We now consider the centrality measures. In Table \ref{tb:str}, we report the in-strength $\textbf{s}^I_k$ and out-strength $\textbf{s}^O_k$ distributions for each network $N_k,\,k = 1, ..., 4$ as well the average values. Notice that the average values for the in-strengths and the out-strengths are the same, due to the property of associativity of the sum.
\begin{table}[h!]
\begin{center}
\begin{tabular}{l|rr|rr|rr|rr}
$V$ & $\textbf{s}^I_1$ & $\textbf{s}^O_1$ & $\textbf{s}^I_2$ & $\textbf{s}^O_2$ & $\textbf{s}^I_3$ & $\textbf{s}^O_3$ & $\textbf{s}^I_4$ & $\textbf{s}^O_4$\\ \hline
a & $396$ & $ 6$ & $396$ & $200$ & $350$ & $200$ & $350$ & $396$ \\
b & $ 8$ & $400$ & $ 8$ & $400$ & $202$ & $100$ & $398$ & $100$ \\
c & $ 4$ & $ 0$ & $392$ & $ 0$ & $398$ & $300$ & $398$ & $300$ \\
d & $ 8$ & $400$ & $ 8$ & $400$ & $100$ & $200$ & $100$ & $200$ \\
e & $396$ & $ 6$ & $396$ & $200$ & $150$ & $400$ & $150$ & $400$ \\ \hline
mean & & $ 162.4 $ & & $ 240 $ & & $ 240 $ & & $ 279.2 $
\end{tabular}
\end{center}
\caption{In-strength $\textbf{s}^I_k$ and out-strength $\textbf{s}^O_k$ for each example $k = 1, \dots, 4$.}
\label{tb:str}
\end{table}

We classify the nodes as big or small with respect to the average values of the corresponding centrality measure.
From Table \ref{tb:str}, it is possible to observe that in network $N_1$, nodes $a$ and $e$ are big receivers but small spreaders. On the contrary, $b$ and $d$ are small receivers and big spreaders and node $c$ is peripheral.
In network $N_2$, the node centrality is the same as the one in network $N_1$, with the exception of node $c$ that becomes a big receiver.
In network $N_3$, nodes of the absorbing class $\{b, d\}$ are peripheral, and the node $a$ is a big receiver but a small spreader. Node $e$ is a small receiver but a big spreader, and $c$ is central.
Centralities in the network $N_4$ differ only for node $a$ that also becomes a big spreader, consequentially a central node. Hence $b$ is affected by such variation and becomes a big receiver.

The homogeneous Markov chains related to the networks of the examples share the same state space $V=\{a,b,c,d,e\}$.
We denote the Markov chain associated to the network $N_k$ by $\mathcal{X}_k$, for $k=1,\dots,4$.
The transition matrix and the initial probability distribution of $\mathcal{X}_k$ can be obtained by the adjacency matrix $\textbf{W}_k$ and will be labelled by $\textbf{P}_k$ and $\pi_{0, k}$. We here list the transition matrices.
$$
\textbf{P}_1=\left(
\begin{array}{lllll}
0 & 0.\bar{3} & 0.\bar{3} & 0.\bar{3} & 0 \\
0.99 & 0 & 0 & 0.01 & 0 \\
0 & 0 & 1 & 0 & 0 \\
0 & 0.01 & 0 & 0 & 0.99 \\
0 & 0.\bar{3} & 0.\bar{3} & 0.\bar{3} & 0 \\
\end{array}
\right),\qquad
\textbf{P}_2=\left(
\begin{array}{lllll}
0 & 0.01 & 0.98 & 0.01 & 0 \\
0.99 & 0 & 0 & 0.01 & 0 \\
0 & 0 & 1 & 0 & 0 \\
0 & 0.01 & 0 & 0 & 0.99 \\
0 & 0.01 & 0.98 & 0.01 & 0 \\
\end{array}
\right)
$$
$$
\textbf{P}_3=\left(
\begin{array}{lllll}
0 & 0.01 & 0.99 & 0 & 0 \\
0 & 0 & 0 & 1 & 0 \\
0.5 & 0 & 0 & 0 & 0.5 \\
0 & 1 & 0 & 0 & 0 \\
0.5 & 0 & 0.5 & 0 & 0 \\
\end{array}
\right),\qquad
\textbf{P}_4=\left(
\begin{array}{lllll}
0 & 0.5 & 0.5 & 0 & 0 \\
0 & 0 & 0 & 1 & 0 \\
0.5 & 0 & 0 & 0 & 0.5 \\
0 & 1 & 0 & 0 & 0 \\
0.5 & 0 & 0.5 & 0 & 0 \\
\end{array}
\right)
$$
The initial probability distributions are proportional to the out-strengths because of Equation \eqref{initial_prob_distr}. They can be written as follows:
\begin{eqnarray*}
\pi_{0,1}^\top &=& [0.008, 0.493, 0, 0.493, 0.007]\\
\pi_{0,2}^\top &=& [0.1\bar{6}, 0.\bar{3}, 0, 0.\bar{3}, 0.1\bar{6}]\\
\pi_{0,3}^\top &=& [0.1\bar{6}, 0.08\bar{3}, 0.25, 0.1\bar{6}, 0.\bar{3}]\\
\pi_{0,4}^\top &=& [0.284, 0.072, 0.215, 0.143, 0.287].
\end{eqnarray*}

The states of the Markov chains $\mathcal{X}$ can be easily classified.
$c$ is an absorbing state for $\mathcal{X}_1$ and $\mathcal{X}_2$ and the set of states $\{b,d\}$ is an absorbing class for the Markov chains $\mathcal{X}_3$ and $\mathcal{X}_4$.
In accordance with the related networks, the probabilities of being absorbed by $c$ and $\{b,d\}$ represent the elements distinguishing $\mathcal{X}_1$ from $\mathcal{X}_2$ and $\mathcal{X}_3$ from $\mathcal{X}_4$, respectively.

Now we will inspect the assortativity as the correlation between the out-strength and the in-strength associated with the states of the mentioned Markov Chains.



\begin{table}[h!]
\begin{center}
\begin{tabular}{r|rrrr}
$h$ & $N_1$ & $N_2$ & $N_3$ & $N_4$ \\ \hline
$1$ & $0.7718$ & $0.0828$ &$ 0.5404$ & $0.5503$ \\
$2$ & $-0.7664$ & $-0.0739$ & $0.5204$ & $-0.0309$ \\
$3$ & $0.6031$ & $0.0507$ & $0.3515$ & $0.2663$ \\
$4$ & $-0.6031$ & $-0.0470$ & $0.5067$ & $-0.0130$ \\
$5$ & $0.5086$ & $0.0361$ & $0.4034$ & $0.1688$ \\
$6$ & $-0.5550$ & $-0.0339$ & $0.4551$ & $-0.0236$ \\
$\cdots$ & $\cdots$ & $\cdots$ & $\cdots$ & $\cdots$ \\
$9995$ & $0$ & $0$ & $0.0098$ & $0.0718$ \\
$9996$ & $0$ & $0$ & $-0.0098$ & $-0.0718$ \\
$9997$ & $0$ & $0$ & $0.0098$ & $0.0718$ \\
$9998$ & $0$ & $0$ & $-0.0098$ & $-0.0718$ \\
$9999$ & $0$ & $0$ & $0.0098$ & $0.0718$ \\
$10000$ & $0$ & $0$ & $-0.0098$ &$-0.0718$
\end{tabular}
\end{center}
\caption{First and last evaluated values of $r^{OI}_h$ of the four illustrative examples.}
\label{tb:toy_res}
\end{table}


\begin{figure}[h!]
\begin{center}
\includegraphics[width = 0.7\textwidth, page=2]{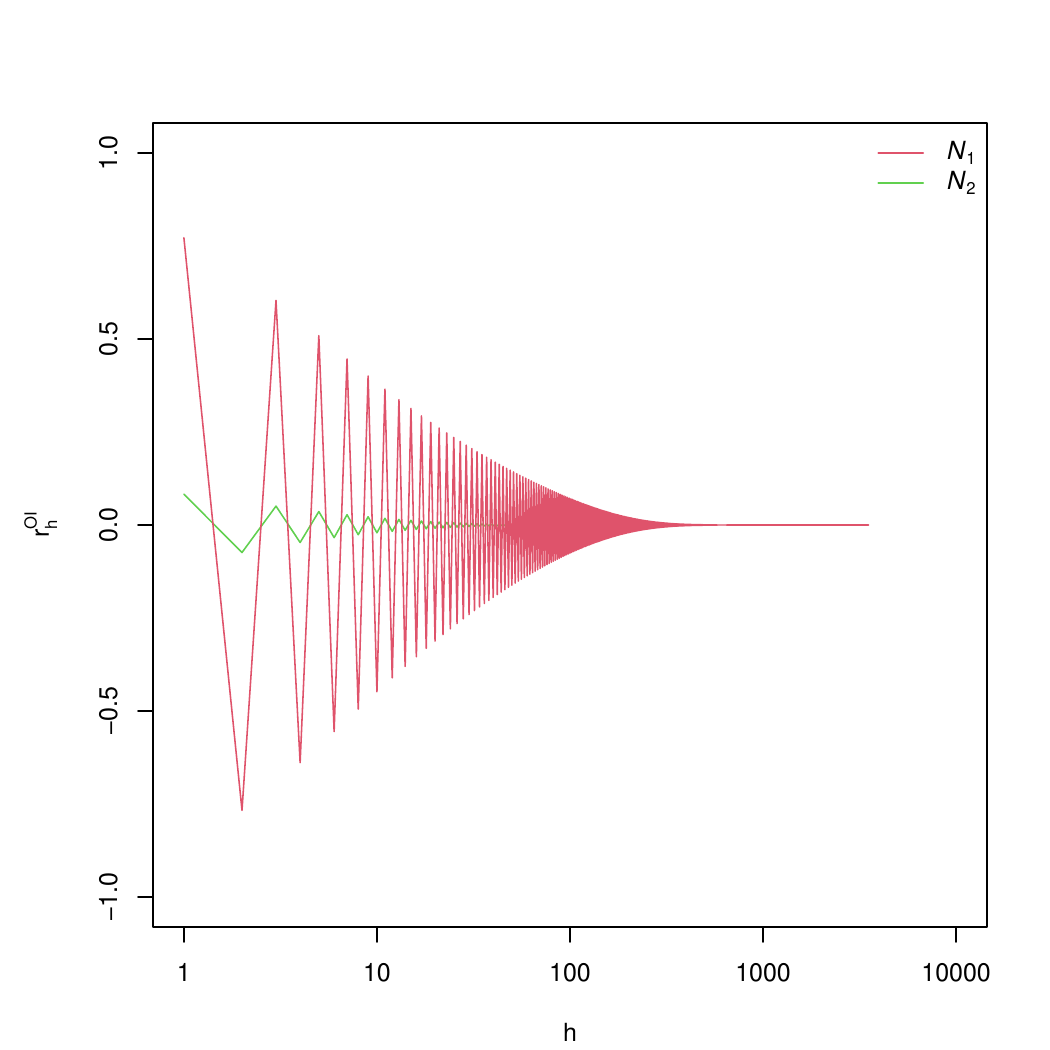}
\end{center}
\caption{$r^{OI}_h$ for increasing values of $h$ of the networks $N_1$ and $N_2$.}
\label{fg:toy_res12}
\end{figure}
Figure \ref{fg:toy_res12} reports the higher order assortativity out-in, $r^{OI}_h$, for the networks $N_1$ and $N_2$. In $N_1$ the probability of falling in the absorbing state $c$ is small; therefore, we can consider its effect as marginal.
Nodes $a$ and $e$ have a small out-strength and high in-strength and in contrast to $b$ and $d$, which have opposite characteristics.
We identify the pair $\{a, e\}$ as group 1 and the pair $\{b, d\}$ as group 2.
At the first step, $h = 1$, the assortativity is due mainly to the correlation between the out-strengths in group 1 with the in-strengths in group 2, and vice-versa.
It can be observed that there is concordance between them and, therefore, positive assortativity.
After two steps, $h=2$, if the Markov Chain starts from group 1, it is highly probable that it jumps to group 2 and back to group 1, generating a disassortative behaviour. It is the same if the Markov Chain starts from group 2.
As a consequence, for increasing values of $h$, this creates the alternation between positive and negative assortativity for odd and even values of $h$, respectively, as confirmed in Table \ref{tb:toy_res}.
The asymptotic zero-assortative behaviour is caused by both the randomness and the presence of the absorbing state $c$.
In network $N_2$, it can be observed the same behaviour with a smaller oscillation. This is due to the increased out-degrees of nodes in group 1 and to the higher probability of falling in the absorbing state $c$ where there is no centrality variability.

Observe that, even if in the sink there is no variability of strength measures, there is not an indeterminate-form in Equation \eqref{eq:higher} because the probability of falling in the sink is less than one.
Moreover in the network $N_1$, the probability of falling in the sink is high; therefore, the asymptotic value is reached immediately. On the contrary, the asymptotic value is reached after a long alternation between assortative and disassortative behaviour in the network $N_2$.

\begin{figure}[h!]
\begin{center}
\includegraphics[width = 0.7\textwidth, page=3]{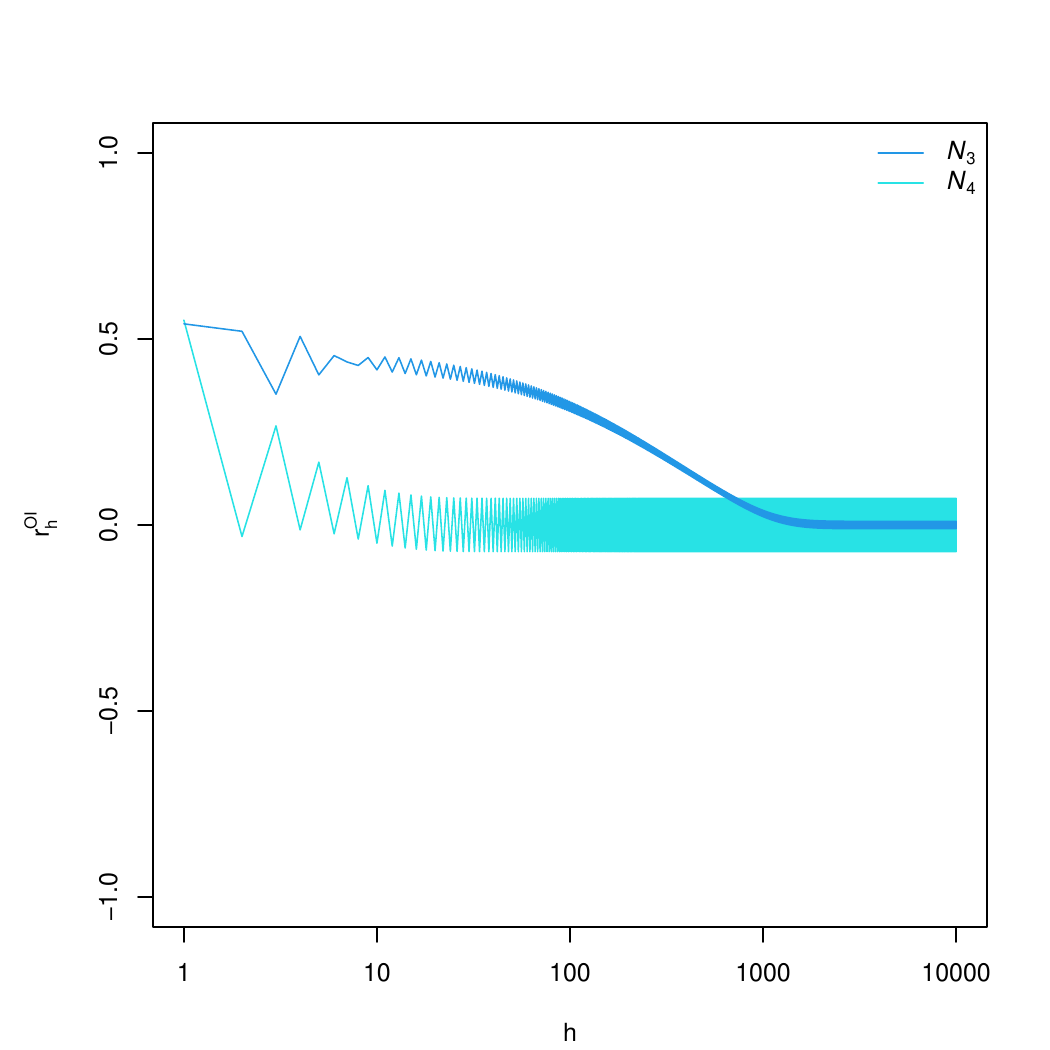}
\end{center}
\caption{$r^{OI}_h$ for increasing values of $h$ of the networks $N_3$ and $N_4$.}

\label{fg:toy_res34}
\end{figure}
Figure \ref{fg:toy_res34} shows the same results for networks $N_3$ and $N_4$.
Such networks are characterized by the absorbing class $\{b, d\}$ and the complementary one $\{a, c, e\}$ connected by the bridge represented by the edge from $a$ to $b$.
Both classes have positive assortativity.
In $N_3$, the bridge has a low weight; therefore, the inter-assortative behaviour dominates.
In network $N_4$, the weight of the bridge is very high, and consequentially the larger oscillations are justified by the differences between the classes.


\section{Conclusions}\label{Sec. Conc}
This paper fills a gap in the complex network literature by introducing a concept of higher order assortativity measure for weighted and directed networks, where the considered nodes' attributes are the directed version of degree and strength centrality. 
Such measures have a clear interpretation in application fields well-linked to direct and weighted networks, such as that of the international trade, studied in the proposed empirical application. Furthermore, we also highlight the strong connection of the introduced assortativity measure and the autocorrelations of suitably defined Markov chains. 

The versatility of the methodological proposal makes it possible to describe the preferential attachment of a wide set of models. 
Some paradigmatic empirical instances might involve the analysis of the roots of the migration flows \citep{ccgmigr} or the structure and systemic risk profile of the interbanking system \citep{bocap15, castellano2021, caseboli, cerqueti2022, cerqueti2021, Lillo15}. Moreover, the readability of the proposed assortativity measure in the context of Markov chains might be efficiently exploited to analyse the properties of some classes of dynamical random systems having the Markovian properties. In this respect, it is important to notice that the decay of the autocorrelation explains the long-term memory properties of the underlying stochastic process \citep[see][]{hurst}. 
Such challenging themes are already in our research agenda.

\bibliographystyle{apa}
\bibliography{assortativity}

\begin{thebibliography}{}

\bibitem[\protect\astroncite{Antonietti et~al.}{2022}]{antonietti2022world}
Antonietti, R., Falbo, P., Fontini, F., Grassi, R., and Rizzini, G. (2022).
\newblock The world trade network: country centrality and the covid-19
  pandemic.
\newblock {\em Applied Network Science}, 7(1):18.

\bibitem[\protect\astroncite{Arcagni et~al.}{2017}]{Arcagni2017}
Arcagni, A., Grassi, R., Stefani, S., and Torriero, A. (2017).
\newblock Higher order assortativity in complex networks.
\newblock {\em European Journal of Operational Research}, 262(2):708--719.

\bibitem[\protect\astroncite{Arcagni et~al.}{2019}]{arcagni2019extending}
Arcagni, A., Grassi, R., Stefani, S., and Torriero, A. (2019).
\newblock Extending assortativity: An application to weighted social networks.
\newblock {\em Journal of Business Research}.

\bibitem[\protect\astroncite{Bang-Jensen and Gutin}{2008}]{Bang2008}
Bang-Jensen, J. and Gutin, G.~Z. (2008).
\newblock {\em Digraphs: theory, algorithms and applications}.
\newblock Springer Science \& Business Media.

\bibitem[\protect\astroncite{Bargigli et~al.}{2015}]{Lillo15}
Bargigli, L., Di~Iasio, G., Infante, L., Lillo, F., and Pierobon, F. (2015).
\newblock The multiplex structure of interbank networks.
\newblock {\em Quantitative Finance}, 15(4):673--691.

\bibitem[\protect\astroncite{Bo and Capponi}{2015}]{bocap15}
Bo, L. and Capponi, A. (2015).
\newblock Systemic risk in interbanking networks.
\newblock {\em SIAM Journal on Financial Mathematics}, 6(1):386--424.

\bibitem[\protect\astroncite{Castellano et~al.}{2021}]{castellano2021}
Castellano, R., Cerqueti, R., Clemente, G.~P., and Grassi, R. (2021).
\newblock An optimization model for minimizing systemic risk.
\newblock {\em Mathematics and Financial Economics}, 15:103--129.

\bibitem[\protect\astroncite{Castiglionesi and Eboli}{2018}]{caseboli}
Castiglionesi, F. and Eboli, M. (2018).
\newblock Liquidity flows in interbank networks.
\newblock {\em Review of Finance}, 22(4):1291--1334.

\bibitem[\protect\astroncite{Cerqueti et~al.}{2022}]{cerqueti2022}
Cerqueti, R., Cinelli, M., Ferraro, G., and Iovanella, A. (2022).
\newblock Financial interbanking networks resilience under shocks propagation.
\newblock {\em Annals of Operations Research},
  https://doi.org/10.1007/s10479-022-04567-w.

\bibitem[\protect\astroncite{Cerqueti et~al.}{2019}]{ccgmigr}
Cerqueti, R., Clemente, G.~P., and Grassi, R. (2019).
\newblock A network-based measure of the socio-economic roots of the migration
  flows.
\newblock {\em Social Indicators Research}, 146:187--204.

\bibitem[\protect\astroncite{Cerqueti et~al.}{2021}]{cerqueti2021}
Cerqueti, R., Clemente, G.~P., and Grassi, R. (2021).
\newblock Systemic risk assessment through high order clustering coefficient.
\newblock {\em Annals of Operations Research}, 299:1165--1187.

\bibitem[\protect\astroncite{Estrada et~al.}{2008}]{Estrada2008}
Estrada, E., Hatano, N., and Gutierrez, A. (2008).
\newblock ``{C}lumpiness'' mixing in complex networks.
\newblock {\em Journal of Statistical Mechanics: Theory and Experiment}, page
  P03008.

\bibitem[\protect\astroncite{Foster et~al.}{2010}]{Foster2010}
Foster, J.~G., Foster, D.~V., Grassberger, P., and Paczuski, M. (2010).
\newblock Edge direction and the structure of networks.
\newblock {\em Proceedings of the National Academy of Sciences},
  107(24):10815--10820.

\bibitem[\protect\astroncite{G{\'o}mez et~al.}{2010}]{gomez}
G{\'o}mez, S., Arenas, A., Borge-Holthoefer, J., Meloni, S., and Moreno, Y.
  (2010).
\newblock Discrete-time markov chain approach to contact-based disease
  spreading in complex networks.
\newblock {\em EPL (Europhysics Letters)}, 89(3):38009.

\bibitem[\protect\astroncite{Gross and Yellen}{2003}]{Gross2003}
Gross, J.~L. and Yellen, J. (2003).
\newblock {\em Handbook of graph theory}.
\newblock CRC press.

\bibitem[\protect\astroncite{Hurst}{1951}]{hurst}
Hurst, H.~E. (1951).
\newblock Long-term storage capacity of reservoirs.
\newblock {\em Transactions of the American society of civil engineers},
  116(1):770--799.

\bibitem[\protect\astroncite{Iannelli et~al.}{2017}]{iannelli}
Iannelli, F., Koher, A., Brockmann, D., H{\"o}vel, P., and Sokolov, I.~M.
  (2017).
\newblock Effective distances for epidemics spreading on complex networks.
\newblock {\em Physical Review E}, 95(1):012313.

\bibitem[\protect\astroncite{Leung and Chau}{2007}]{Leung2007}
Leung, C. and Chau, H. (2007).
\newblock Weighted assortative and disassortative networks model.
\newblock {\em Physica A: Statistical Mechanics and its Applications},
  378(2):591--602.

\bibitem[\protect\astroncite{Mayo et~al.}{2015}]{Mayo2015}
Mayo, M., Abdelzaher, A., and Ghosh, P. (2015).
\newblock Long-range degree correlations in complex networks.
\newblock {\em Computational Social Networks}, 2(1):1--13.

\bibitem[\protect\astroncite{Meghanathan}{2016}]{Meghanathan2016}
Meghanathan, N. (2016).
\newblock Assortativity analysis of real-world network graphs based on
  centrality metrics.
\newblock {\em Computer and Information Science}, 9(3):7--25.

\bibitem[\protect\astroncite{Newman}{2010}]{Newman2010_DOPPIO}
Newman, M. (2010).
\newblock {\em Networks: an introduction}.
\newblock Oxford university press.

\bibitem[\protect\astroncite{Newman}{2002}]{Newman2002}
Newman, M. E.~J. (2002).
\newblock Assortative {Mixing} in {Networks}.
\newblock {\em Physical Review Letters}, 89(20):208701.

\bibitem[\protect\astroncite{Norris et~al.}{1998}]{norris}
Norris, J.~R., Norris, J.~R., and Norris, J.~R. (1998).
\newblock {\em Markov chains}.
\newblock Number~2 in Cambridge Series in Statistical and Probabilistic
  Mathematics. Cambridge University Press.

\bibitem[\protect\astroncite{Piraveenan et~al.}{2010}]{Piraveenan2010}
Piraveenan, M., Prokopenko, M., and Zomaya, A. (2010).
\newblock Assortative mixing in directed biological networks.
\newblock {\em IEEE/ACM Transactions on computational biology and
  bioinformatics}, 9(1):66--78.

\bibitem[\protect\astroncite{{UN COMTRADE}}{2022}]{UNComtrade}
{UN COMTRADE} (2022).
\newblock International trade statistics database.
\newblock data retrieved from International Trade Statistics Database,
  \url{https://comtrade.un.org/}.

\bibitem[\protect\astroncite{Van~Mieghem et~al.}{2010}]{vanmieghen2010}
Van~Mieghem, P., Wang, H., Ge, X., Tang, S., and Kuipers, F.~A. (2010).
\newblock Influence of assortativity and degree-preserving rewiring on the
  spectra of networks.
\newblock {\em The European Physical Journal B}, 76(4):643--652.

\bibitem[\protect\astroncite{Yuan et~al.}{2021}]{Yuan2021}
Yuan, Y., Yan, J., and Zhang, P. (2021).
\newblock Assortativity measures for weighted and directed networks.
\newblock {\em Journal of Complex Networks}, 9(2):cnab017.

\end{thebibliography}

\end{document}